\documentclass[aps,prd,superscriptaddress,showkeys,showpacs,twocolumn,a4paper]{revtex4-1}
\usepackage{ulem}
\usepackage{color}
\usepackage{graphicx}
\usepackage{appendix}
\usepackage{epsfig}
\usepackage[colorlinks=true,linkcolor= red, citecolor = cyan, urlcolor = teal ]{hyperref}
\usepackage{epstopdf}
\usepackage{amsmath}
\usepackage{subfig}
\usepackage{comment}
\usepackage{float}
\usepackage{float}
\usepackage[dvipsnames]{xcolor}
\newcommand{\be}{\begin{equation}}
\newcommand{\ee}{\end{equation}}
\newcommand{\ba}{\begin{eqnarray}}
\newcommand{\ea}{\end{eqnarray}}

\begin{document}
\title{Heavy quark radiation in an anisotropic hot QCD medium }
\author{Jai Prakash}
\email{jai183212001@iitgoa.ac.in}
\affiliation{School of Physical Science, Indian Institute of Technology Goa, Ponda-403401, Goa, India}
\author{Vinod Chandra}
\email{vchandra@iitgn.ac.in}
\affiliation{Indian Institute of Technology Gandhinagar, Gandhinagar-382355, Gujarat, India}
\author{Santosh K. Das}
\email{santosh@iitgoa.ac.in}
\affiliation{School of Physical Science, Indian Institute of Technology Goa, Ponda-403401, Goa, India}


\begin{abstract}
The impact of momentum anisotropy { of the medium} on the heavy quarks (HQs) dynamics has been investigated in a hot QCD medium while considering both collisional and radiative processes within the ambit of the Fokker–Planck approach. The relative orientation of the HQs motion (momentum vector) with respect to the direction of anisotropy is responsible for the character of transport coefficients. Therefore, the drag and diffusion coefficients of the HQs are decomposed, respectively, into two and four components by considering a general tensor basis. The transport coefficients of the HQs are decomposed into two (drag coefficients) and four (diffusion coefficients)  components by assuming a general tensor basis. Each component of the drag and diffusion coefficient of the HQs has been analyzed in detail. It is observed that the anisotropy has a significant impact on the transport coefficients of the HQ for both the collisional and the radiational processes. The nuclear suppression factor, $R_{AA}$, has been computed considering the anisotropic medium. It is observed that the momentum anisotropy affects the $R_{AA}$ of the HQs significantly in both elastic and inelastic cases.
\end{abstract}



\maketitle

 \section{Introduction}
 \label{intro}
There are strong indications that the hot QCD matter, commonly termed as Quark-Gluon Plasma (QGP), has already been observed in ultra-relativistic heavy-ion collisions (HICs) at Relativistic Heavy-ion Collider (RHIC) and the Large Hadron Collider (LHC)~\cite{STAR:2006vcp,Adams:2005dq,PHENIX:2004vcz,ALICE:2010khr,BRAHMS:2005gow}. The QGP turned out to be short-lived  with an expected lifetime of a few fm/c ($4–5$ fm/c at RHIC and $10–12$ fm/c at LHC)~ \cite{vanHees:2004gq,Rapp:2009my}.

On the other hand, understanding the properties of the QGP through the HQs (charm and beauty) dynamics is a field of high interest as they are considered to be one of the most prominent probes~  \cite{Das:2016llg,Song:2015sfa,Andronic:2015wma,Dong:2019unq,Cao:2018ews,Rapp:2018qla,Prino:2016cni,Aarts:2016hap,Uphoff:2011ad,GolamMustafa:1997id,Plumari:2017ntm,Gossiaux:2008jv, Das:2017dsh,Das:2022lqh} since they are produced at a very early stage in HICs and due to having a huge mass, their thermalization time is quite larger than the constituents particles of the medium. Consequently, HQs can act as an excellent probe to witness the entire space-time evolution of the QGP medium and acts as non-equilibrium degrees of freedom in the equilibrated QGP. In several studies, the associated experimental observables, such as heavy-quark nuclear suppression factor, $R_{AA}$, have been discussed within the framework of the Langevin dynamics \cite{Rapp:2009my,vanHees:2007me,Das:2015ana,Ruggieri:2022kxv}. The inclusion of momentum anisotropy in these analyses is another crucial aspect in the study of the HQs dynamics. 

The momentum anisotropy { of the medium} may induce instabilities in the Yang-Mills fields that may lead to turbulence in the plasma ~\cite{Mrowczynski:1993qm,Randrup:2003cw,Chandra:2012qq,Chandra:2011bu} and could help in understanding the evolution of the QGP medium. The presence of momentum anisotropy causes the chromo-Weibel instability in the hot QCD medium that affects the transport coefficients of the HQs and modifies the energy loss of the HQs in the QCD medium. This has been studied for collisional processes in Ref.~\cite{Chandra:2015gma}  and later extended for radiational processes in Ref.~\cite{Prakash:2021lwt}. In the HICs, the fast evolution of the QGP medium leads to an imbalance in the longitudinal and transverse expansions and causes momentum anisotropy in the medium. Therefore, it is interesting to investigate the impact of momentum anisotropy on the properties of the QGP. Earlier, the anisotropic aspects have been studied in the context of electromagnetic probes ~\cite{Romatschke:2003ms,kasmaei2020photon,PhysRevD.99.034015}, gluon self-energy~\cite{Ghosh:2020sng}, heavy quark and jet in the pre-equilibrium phase ~\cite{Boguslavski:2023fdm,Boguslavski:2023alu}, to evaluate the impact of shear and bulk viscous coefficients of the HQs transport in the medium ~\cite{Kurian:2020kct,Song:2019cqz,Kurian:2020orp}. 
The effect of momentum anisotropy has been observed for the quarkonium states \cite{Dumitru:2009ni}.

In this work, our focus is to study the effect of anisotropy in the hot QCD medium via the transport coefficient of the HQs considering collisional and radiative processes (as gluon bremsstrahlung) \cite{Shaikh:2021lka,Mustafa:2004dr,Sarkar:2018erq,Braaten:1991we,Cao:2013ita}. Ref. \cite{Mazumder:2013oaa,Prakash:2023hfj} has already performed a comparative study on the transport coefficient for elastic and inelastic processes in the isotropic medium. The properties of the medium will be studied through the transport coefficients; namely, the drag and the diffusion coefficients of the HQs within the framework of stochastic Langevin dynamics \cite{Das:2013kea} where the transport coefficients are needed as input.  

In this manuscript, a general framework is adopted to evaluate the HQs motion in the momentum anisotropic QGP medium. Based on the relative direction of anisotropy and the HQs momentum, the HQs drag force decomposed into two components and that of momentum diffusion tensor into four components using a tensor basis ~\cite{Kumar:2021goi}. The effect of anisotropy enters through the non-equilibrium distribution of the constituent particle of the hot QCD medium and will be seen through the transport coefficients. We further have studied the $R_{AA}$ of the HQs in the medium, considering both the radiative and collisional cases. The major focus is on the impact of anisotropy on the HQs dynamics in inelastic processes in the medium. 

The article is organized as follows. Section~ \ref{form}, is dedicated to the dynamics of the HQ through the transport coefficient for the collisional and radiative processes. Here, the transport coefficients are shown to decompose in several components in the anisotropic medium. In section~ \ref{res}, we present the results of the current analyses. Finally, the work is concluded in section~ \ref{con}.

\noindent {\bf Notations and conventions}:
The subscript $k$ in the manuscript represents the particle species, $i.e.$, $k=(g, {q})$, where $g$ and ${q}$  denote gluons and light quarks, respectively. The energy of the HQs is denoted by $E_p$ and can be written as $E_p =\sqrt{{\bf{|p|^2}} + M_{HQ}^2}$, where $M_{HQ}$ is the mass of the HQs. The energy of a parton is denoted by $E_q$, where the constituent particles are massless.

 \section{Formalism}
 \label{form}
The dynamics of the HQs in the QGP medium has been carried within the framework of the non-equilibrium Boltzmann equation given as ~\cite{Svetitsky:1987gq},
 \begin{equation}\label{11}
  p^{\mu}\partial_{\mu}f_{HQs}=\bigg(\dfrac{\partial f_{HQs}}{\partial t}\bigg)_{\text{int}},
\end{equation}
where $f_{HQs}$ represents the momentum distribution of the HQs. We follow the assumption of soft momentum transfer as a result the relativistic non-linear Boltzmann equation will be converted to  the Fokker-Plank  equation \cite{Mazumder:2011nj} as follows, 
\begin{align}\label{14}
  	\frac{\partial f_{HQs}}{\partial t}=\frac{\partial}{\partial p_i}\left[\gamma_i({\bf p})f_{HQs}+\frac{\partial}{\partial p_j}\Big(D_{i j}({\bf p})f_{HQs}\Big)\right],
  	\end{align}
 where $\gamma_i$ is the drag coefficients and  $D_{ij}$ the diffusion coefficients of the HQs. The transport coefficients govern the dynamics of the HQ in the QGP medium, where the HQs are treated as Brownian particles. Below, we discuss collisional and radiative aspects of HQ transport in the presence of momentum anisotropy. 
 
\subsection{\bf Collisional processes} 
When the HQs propagate through the medium, they go through collisional interactions with the medium particles. Two-component (HQs and medium particles) interactions in collisional processes are,  $HQs(P)+l(Q)\rightarrow HQs(P^{'})+l(Q^{'})$, where $P, Q$  are four-momentum before the collision, and $P', Q'$ are four-momentum after the collision~\cite{Svetitsky:1987gq}, and $l$ stand for light quarks, anti-quarks, and gluons.  

Next, the transport coefficients of the HQ can be expressed in terms of the momentum of the partons as ~\cite{Svetitsky:1987gq},
\begin{align}\label{15}
    \gamma_i=&\frac{1}{2E_p}\int{\frac{d^3{\bf q}}{(2\pi)^32E_q}}\int{\frac{d^3{\bf q}'}{(2\pi)^32E_{q'}}}\int{\frac{d^3{\bf p}'}{(2\pi)^32E_{p'}}}\frac{1}{\gamma_{HQ}}\nonumber\\ &\times\sum|\mathcal{M}_{2\rightarrow 2}|^2(2\pi)^4\delta^4 (P+Q-P'-Q') f_{k}({\bf{q}})\nonumber\\ &\times\Big(1+a_k f_{k}({\bf{q'}})\Big)\Big[({\bf p}-{\bf p}')_i\Big]\nonumber\\
	&=\langle\langle({\bf p}-{\bf p}')_i\rangle\rangle,
\end{align}

 and, 
\begin{align}\label{16}
    D_{ij}=&\frac{1}{2E_p}\int{\frac{d^3{\bf q}}{(2\pi)^32E_q}}\int{\frac{d^3{\bf q}'}{(2\pi)^32E_{q'}}}\int{\frac{d^3{\bf p}'}{(2\pi)^32E_{p'}}}\frac{1}{\gamma_{HQ}}\nonumber\\ &\times\sum|\mathcal{M}_{2\rightarrow 2}|^2(2\pi)^4\delta^4 (P+Q-P'-Q') f_{k}({\bf{q}})\nonumber\\ &\times\Big(1+a_k f_{k}({\bf{q'}})\Big)\frac{1}{2}\Big[({\bf p}-{\bf p}')_i({\bf p}-{\bf p}')_j\Big]\nonumber\\
	&=\frac{1}{2}\langle\langle({\bf p}-{\bf p}')_i({\bf p}-{\bf p}')_j\rangle\rangle,
\end{align}
where the energy of partons, $E_q$, and the momentum, $\bf{q}$ correspond to those  before collision whereas, $E_q'$, and $\bf{q'}$ represents those after the collisions with the HQs. The conservation of energy-momentum is taken care of by the delta function. The temperature dependency of transport coefficients enters through the phase space distribution function, $f_{k}({\bf{q'}})$ of medium particles. The scattering matrix element, $\sum|\mathcal{M}_{2\rightarrow 2}|^2$ sustain the information of the HQs interaction with the partons of the hot QCD medium.  The alternative realization of $\gamma_i$ is the thermal average of the momentum transfer of the HQs, while { $D_{ij}$} is the thermal average of the square of the momentum transfer of the HQs. For the isotropic case,
the momentum dependency of $\gamma_i$ and { $D_{ij}$} can be noted as follow,
\begin{align}\label{17}
 &\gamma=\langle\langle 1 \rangle\rangle - \frac{\langle\langle {\bf{p.p'} \rangle\rangle}}{p^2}.
\end{align}
The diffusion coefficients contain two components  in the isotropic medium, the transverse part, $(D_{0})$ and the longitudinal part $(D_{1})$ given as,
\begin{align}\label{19}
&D_{0}= \frac{1}{4}\left[\langle\langle p'^{2} \rangle\rangle-\frac{\langle\langle ({\bf{p'.p}})^2\rangle\rangle}{p^2} \right],\\ 
&D_{1}= \frac{1}{2}\left[\frac{\langle\langle ({\bf{p'.p})}^2\rangle\rangle}{p^2} -2\langle\langle ({\bf{p'.p})}\rangle\rangle +p^2 \langle\langle 1 \rangle\rangle\right]\label{19.1}.
 \end{align}
 With the relevant choice of $F({\bf p})$ in the center of the mass frame, one can write the average of a function, $\langle \langle F(\textbf{p})\rangle\rangle$ for the collisional processes in an isotropic medium as, 
 \begin{align}
    \label{avergef}\langle \langle F(\textbf{p})\rangle\rangle&= \frac{1}{(512\pi^4)\textit{E}_p\gamma_{HQ}}\int_0^{\infty}qdq\left(\frac{s-m^2_{HQ}}{s}\right)\times\nonumber\\ &\int_0^{\pi}d\chi \sin\chi\int_0^{\pi}d\theta_{cm}\sin\theta_{cm}\sum|\mathcal{M}_{2\rightarrow2}|^2
 \times\nonumber\\ &\int_0^{2\pi} d\phi_{cm}\left[\left(1+a_kf_k^0(\textbf{q}^{\prime})\right)f_k^0(\textbf{q}) \right]F(\textbf{p}),
  \end{align}
 where $a_k=-1$, corresponds to Fermi suppression and $a_k=1$ for Bose enhancement regarding the final state phase space for the quarks and gluons, respectively. Here, $s$ is the Mandelstam variable. {The infrared divergence appears in the gluonic propagator within the t-channel diagrams is screened by the  Debye mass, $m_{D}$.}
 Next, we shall discuss the incorporation of anisotropy in the current analyses.  

\subsection{Aspect of anisotropic medium}
 The momentum anisotropy appears in the picture due to the fast expansion of the hot QCD medium, and it is captured through the non-equilibrium distribution function of the medium particles. The general form of momentum anisotropic distribution can  be written as follow, \cite{Romatschke:2003ms,PhysRevD.73.125004},
\begin{align}
 \label{13}&f_k^{(\text{aniso})}(\textbf{q})= \sqrt{1+\xi} \hspace{0.05cm} f_k^0\left(\sqrt{q^2 + \xi(\textbf{q} \hspace{0.05cm} \mathord{\cdot} \hspace{0.05cm} \textbf{n})^2}\right),
\end{align}
where $\xi$ is the anisotropic variable that determines the squeezing ($\xi>0$) or stretching ($-1<\xi<0$) of the momentum distribution along the direction $\bf{n}$, where $\bf{n}$ is the unit vector that defines the direction of momentum anisotropy in the medium. {Inclusion of the normalization factor $\sqrt{1+\xi}$ ensures that the total particle number remains constant for both the anisotropic and isotropic distribution functions~\cite{Schenke:2006xu}.} This formalism has been pursued under the weak isotropic limit, $\xi\ll 1$. The modified distribution function, $f_k^{(\text{aniso})}$  contains isotropic ($f_k^0$) distribution along with the correction term ($\delta f_k$), {\it i,e.,} $f_k^{(\text{aniso})} =f_k^0+\delta f_k,$  as describe in Ref. \cite{Srivastava:2015via}
\begin{align}
\label{deltaf}&\delta f_k= -\frac{\xi}{2\textit{E}_q\textit{T}}(\textbf{q} \hspace{0.05cm} \mathord{\cdot} \hspace{0.05cm} \textbf{n})^2 \hspace{0.05cm} (f_k^0)^2\textit{exp}\left(\frac{\textit{E}_q}{\textit{T}}\right),
\end{align}
where $T$ is the temperature of the thermal bath, and $E_q$ is the energy of the medium particle. The drag coefficient decomposes in the orthogonal basis for the anisotropic medium as \cite{Kumar:2021goi},

\begin{align}
\label{drag_deco}&\gamma_i =p_i\gamma_0^{(\text{aniso})}+\Tilde{n}_i\gamma_1^{(\text{aniso})}.
 \end{align}
 
 The two components of the drag coefficient in the  anisotropic medium can be expressed  as follow,
\begin{align}
 \label{drag_0}&\gamma_0^{(\text{aniso})}=p_i\gamma_i/p^2 = \langle \langle 1\rangle\rangle-\frac{\langle \langle \textbf{p} \hspace{0.05cm} \mathord{\cdot} \hspace{0.05cm} 
 \textbf{p}^{\prime}\rangle\rangle}{p^2},
 \\
 \label{drag_1}&\gamma_1^{(\text{aniso})}=\Tilde{n}_i\gamma_i/\Tilde{n}^2 = -\frac{1}{\Tilde{n}^2}\langle \langle \Tilde{\textbf{n}} \hspace{0.05cm} \mathord{\cdot} \hspace{0.05cm} 
 \textbf{p}^{\prime}\rangle\rangle,
 \end{align}
 where, $\quad\Tilde{n}^2 = 1-\frac{(\textbf{p} \hspace{0.05cm} \mathord{\cdot} \hspace{0.05cm} \Hat{\textbf{n}})^2}{p^2} = 1- \cos^2\theta$ and angle, $\theta$ is between the HQs momentum, and the anisotropy vector of the medium.
 The additional component of drag coefficient, $\gamma_1^{(\text{aniso})}$
sustain the information of relative orientation of momentum anisotropy and the HQs motion. In this case,
the average of $F(\textbf{p})$ for the collisional processes in the COM frame can be written by using Eq.~\eqref{avergef},
 
\begin{align}
    \label{averagedf}\langle \langle &F(\textbf{p})\rangle\rangle_{ca}= \frac{1}{(1024\pi^5)\textit{E}_p\gamma_{HQ}}\int_0^{\infty}qdq\left(\frac{s-m^2_{HQ}}{s}\right)\times\nonumber\\ &\int_0^{\pi}d\chi \sin\chi\int_0^{2\pi}d\phi \int_0^{\pi}d\theta_{cm}\sin\theta_{cm}\sum|\mathcal{M}_{2\rightarrow2}|^2
 \times\nonumber\\ &\int_0^{2\pi} d\phi_{cm}\left[\delta f_k(\textbf{q})\left(1+a_kf_k^0(\textbf{q}^{\prime})\right) + a_kf_k^0(\textbf{q})\delta f_k(\textbf{q}^{\prime})\right]F(\textbf{p}),
  \end{align}
the transport coefficients of the HQ for the collisional processes can be described as a combination of isotropic in Eq.~\eqref{avergef} and anisotropic medium in Eq.~\eqref{averagedf}, as follows,
\begin{align}
\label{trans_coll}
X_c = X_{c0} +X_{ca},
\end{align}
the component of drag coefficient, $\gamma_0^{(\text{aniso})}$ contains the anisotropic correction that enters through the non-equilibrium distributions,
 \begin{align}
 \label{20}&\gamma_0^{(\text{aniso})}= \gamma_0+\delta \gamma_0,
 \end{align}
 in the anisotropic medium, the diffusion coefficient also decomposes into four components, as also shown in Ref.~\cite{Romatschke:2003ms}, given as,
 \begin{align}
     \label{diff}D_{ij}=&\left(\delta_{ij}-\frac{p_ip_j}{p^2}\right)D_0^{(\text{aniso})}+\frac{p_ip_j}{p^2}D_1^{(\text{aniso})}+\frac{\Tilde{n}_i\Tilde{n}_j}{\Tilde{n}^2}D_2^{(\text{aniso})} \nonumber\\ & +\left(p^i\Tilde{n}^j + p^j\Tilde{n}^i \right)D_3^{(\text{aniso})}.
  \end{align}
Extracting all the diffusion coefficient components by taking proper projection from the Eq.~\eqref{diff} ~\cite{Kumar:2021goi},
 \begin{align}
  \label{22}D_0^{(\text{aniso})}&=\left[\left(\delta_{ij}-\frac{p_ip_j}{p^2}\right)-\frac{\Tilde{n}_i\Tilde{n}_j}{\Tilde{n}^2}\right]D_{ij},\nonumber\\ &=\frac{1}{2}\left[\langle \langle p^{\prime 2}\rangle\rangle -\frac{\langle \langle (\textbf{p}^{\prime} \hspace{0.05cm} \mathord{\cdot} \hspace{0.05cm} \textbf{p} )^2\rangle\rangle}{p^2}-\frac{\langle \langle (\textbf{p}^{\prime} \hspace{0.05cm} \mathord{\cdot} \hspace{0.05cm} \Tilde{\textbf{n}} )^2\rangle\rangle}{\Tilde{n}^2}\right].
    \end{align}
    \begin{align}
    \label{23}D_1^{(\text{aniso})}&= \frac{p_ip_j}{p^2}D_{ij},\nonumber\\ &= \frac{1}{2}\left[\frac{\langle \langle (\textbf{p}^{\prime} \hspace{0.05cm} \mathord{\cdot} \hspace{0.05cm} \textbf{p} )^2\rangle\rangle}{p^2}-2\langle \langle (\textbf{p}^{\prime} \hspace{0.05cm} \mathord{\cdot} \hspace{0.05cm} \textbf{p} )\rangle\rangle -p^2{\langle \langle 1\rangle\rangle} \right],
   \end{align}
   \begin{align}
       \label{24}D_2^{(\text{aniso})}&= \left[\frac{2\Tilde{n}_i\Tilde{n}_j}{\Tilde{n}^2}-\left(\delta_{ij}-\frac{p_ip_j}{p^2}\right)\right]D_{ij},\nonumber\\ &=\frac{1}{2}\left[-\langle \langle p^{\prime 2} \rangle\rangle+\frac{\langle \langle (\textbf{p}^{\prime} \hspace{0.05cm} \mathord{\cdot} \hspace{0.05cm} \textbf{p} )^2\rangle\rangle}{p^2} +\frac{2\langle \langle (\textbf{p}^{\prime} \hspace{0.05cm} \mathord{\cdot} \hspace{0.05cm} \Tilde{\textbf{n}} )^2\rangle\rangle}{\Tilde{n}^2}\right],
 \end{align}
 \begin{align}
   \label{25}D_3^{(\text{aniso})}&= \frac{1}{2p^2\Tilde{n}^2}\left(p^i\Tilde{n}^j + p^j\Tilde{n}^i \right)D_{ij},\nonumber\\ &=\frac{1}{2p^2\Tilde{n}^2}\left[-p^2\langle \langle (\textbf{p}^{\prime} \hspace{0.05cm} \mathord{\cdot} \hspace{0.05cm} \Tilde{\textbf{n}} )\rangle\rangle+ \langle \langle(\textbf{p}^{\prime} \hspace{0.05cm} \mathord{\cdot} \hspace{0.05cm} {\textbf{p}} ) (\textbf{p}^{\prime} \hspace{0.05cm} \mathord{\cdot} \hspace{0.05cm} \Tilde{\textbf{n}} )\rangle\rangle\right].
 \end{align}
The term $\langle \langle (\textbf{p}^{\prime} \hspace{0.05cm} \mathord{\cdot} \hspace{0.05cm} \Tilde{\textbf{n}} )\rangle\rangle $ contains the information of relative angle of the HQs momentum to the anisotropy vector. One can write $\langle \langle (\textbf{p}^{\prime} \hspace{0.05cm} \mathord{\cdot} \hspace{0.05cm} \Tilde{\textbf{n}} )\rangle\rangle $ in the COM frame, where, we have taken $\textbf{n} = (\sin{\theta},0,\cos{\theta})$. In the anisotropic medium, the momentum of partons, $\textbf{q}$ can be decomposed as $( q\sin\chi\cos\phi,q\sin\chi\sin\phi,q\cos \chi)$. In this analysis, the momentum of the HQs \textbf{p} is considered in the z-direction, such as,

 \begin{align}
     \label{26}&\textbf{p} \hspace{0.05cm} \mathord{\cdot} \hspace{0.05cm} {\textbf{q}}= pq\cos\chi,
   \\
  \label{27}&\textbf{p} \hspace{0.05cm} \mathord{\cdot} \hspace{0.05cm} {\textbf{n}}= p\cos\theta,
   \\
  \label{28}&\textbf{q} \hspace{0.05cm} \mathord{\cdot} \hspace{0.05cm} {\textbf{n}}= q\sin\chi\cos\phi\sin\theta +q\cos \chi\cos\theta,
   \end{align}

further, we can define,
  \begin{align}
      \label{29}&\langle \langle  \Tilde{\textbf{n}}\hspace{0.05cm} \mathord{\cdot} \hspace{0.05cm} \textbf{p}^{\prime} \rangle\rangle =\langle \langle \textbf{n} \hspace{0.05cm} \mathord{\cdot} \hspace{0.05cm} \textbf{p}^{\prime} \rangle\rangle-
    \langle \langle \textbf{p} \hspace{0.05cm} \mathord{\cdot} \hspace{0.05cm} \textbf{p}^{\prime} \rangle\rangle\frac{\cos\theta}{p},
  \end{align}
  where angle $\chi$ is involved in the interaction of the HQs and medium constituents, $\theta_{cm}$ and $\phi_{cm}$ are zenith and azimuthal angles, respectively,  in the COM frame. The momentum of the HQs, in the Lorentz transformation that relates the laboratory frame and COM frame can be written as,
  \begin{align}
      \label{30}&\textbf{p}^{\prime}  = \gamma_{cm}\left( \Hat{\textbf{p}}^{\prime}_{cm} +
  \textbf{v}_{cm} \Hat{E}^{\prime}_{cm} \right),
    \end{align}
 here, we can write the COM velocity ($\textbf{v}_{cm}$) and $\quad \gamma_{cm}$ as, 
  $$\quad \gamma_{cm}= \frac{E_p +E_q}{\sqrt{s}},\\
  \quad \textbf{v}_{cm}= \frac{\textbf{p}+\textbf{q}}{E_p +E_q},$$
  accordingly, the HQs momentum in the COM frame can be decomposed as follows, 

  \begin{align}
\label{31}\Hat{\textbf{p}}^{\prime}_{cm} &=\Hat{p}_{cm}(\cos\theta_{cm}\Hat{\textbf{x}}_{cm} +\sin\theta_{cm}\sin\phi_{cm}\Hat{\textbf{y}}_{cm}\nonumber\\ &+\sin\theta_{cm}\cos\phi_{cm}\Hat{\textbf{z}}_{cm} ),
     \end{align}
where  the momentum and energy of the HQs in the COM frame \cite{Svetitsky:1987gq} have
the following forms:
      $$\quad \Hat{p}_{cm} = \frac{s-m_{HQ}^2}{2\sqrt{s}},$$ $$ \quad \Hat{E}_{cm} = \sqrt{\Hat{p}^2_{cm} +m_{HQ}^2}.$$\\
As, $\textbf{x}_{cm},\textbf{y}_{cm},\textbf{z}_{cm}$ defined in the reference \cite{Svetitsky:1987gq} implementing here for the anisotropic case taken from ref. \cite{Kumar:2021goi}

\begin{align}
    \label{32v} \Tilde{\textbf{n}} \mathord{\cdot}  \textbf{p}^{\prime} &=\frac{\gamma_{cm}}{1+\gamma^2_{cm}v^2_{cm}}\Big[\Hat{p}_{cm} \Big(\cos\theta_{cm}(\Hat{\textbf{x}}_{cm} \mathord{\cdot}  \textbf{n})+\sin\theta_{cm}\nonumber\\
  & \times\sin\phi_{cm}(\Hat{\textbf{y}}_{cm}\mathord{\cdot}  \textbf{n})+ \sin\theta_{cm}\cos\phi_{cm}(\Hat{\textbf{z}}_{cm} \mathord{\cdot}  \textbf{n})\Big)+
  \nonumber \\
  &\gamma_{cm}E_p^{\prime} \frac{(p\cos\theta + q\cos\chi \cos\theta + q\sin\chi\cos\phi\sin\theta)}{E_p+E_q} \Big]\nonumber\\
  &-\frac{\gamma_{cm}}{1+\gamma^2_{cm}v^2_{cm}} \frac{\cos\theta}{p}\Big[\Hat{p}_{cm}\Big(\cos\theta_{cm}(\Hat{\textbf{x}}_{cm}\mathord{\cdot}  \textbf{p})+\nonumber\\
  &\sin\theta_{cm}\sin\phi_{cm}(\Hat{\textbf{y}}_{cm} \mathord{\cdot}  \textbf{p}) \Big)+\gamma_{cm}E_p^{\prime}\frac{(p^2+pq\cos\chi)}{E_p+E_q}\Big], 
 \end{align}
and,
\begin{align}
  \label{32b} {\textbf{p}}\hspace{0.05cm} \mathord{\cdot} \hspace{0.05cm} \textbf{p}^{\prime} &=\frac{\gamma_{cm}}{1+\gamma^2_{cm}v^2_{cm}}\Big[\Hat{p}_{cm}(\cos\theta_{cm}(\Hat{\textbf{x}}_{cm}\hspace{0.05cm} \mathord{\cdot} \hspace{0.01cm} \textbf{p})\nonumber\\
  &+\sin\theta_{cm}\sin\phi_{cm}(\Hat{\textbf{y}}_{cm}\hspace{0.05cm} \mathord{\cdot} \hspace{0.05cm} \textbf{p}))+\gamma_{cm}E_p^{\prime}\frac{(p^2+pq\cos\chi)}{E_p+E_q}\Big],\nonumber\\
  &= E_pE_p^{\prime} - \Hat{E}^2_{cm} + \Hat{p}^2_{cm}\cos\theta_{cm}.
\end{align}

The corresponding projections of the HQs momentum and the anisotropy vector in the medium with the COM axis are depicted in apendix.~\ref{apendix}.
\subsection{\bf{Radiative processes}}
The HQs may radiate gluon while propagating through the medium, and hence, the radiative processes also contribute to the HQ transport coefficients along with the collision processes of the HQs with the medium constituents particles in the hot QCD medium.
For the radiative process $2 \rightarrow 3$ as follows, $HQs(P)+l(Q)\rightarrow HQs(P^{'})+l(Q^{'})+g(K_5)$, where $K_5=(E_5, k_{\perp}, k_z)$ is the four-momentum of the emitted gluons from the moving the HQs within the medium. The contribution of the radiative processes to the HQ transport coefficient can be written in terms of collisional processes. The transport coefficient for the radiative processes in isotropic medium ~\cite{Mazumder:2013oaa} is given as (see also~\cite{Prakash:2021lwt,Shaikh:2021lka,Liu:2020dlt,Song:2022wil}),

\begin{align}\label{rad}
 X_{r0}&=X_{c0}\times\int\frac{d^3{\bf k}_5}{(2\pi)^32E_5}12g^2\frac{1}{k^2_\perp}\left(1+\frac{M_{HQ}^2}{s}e^{2\eta}\right)^{-2}\nonumber\\
    &\times\Big(1+ \hat{f}(E_5)\Big)\Theta_1(\tau-\tau_F)\Theta_2(E_p-E_5), 
\end{align}
and similarly for the anisotropic medium \cite{Prakash:2021lwt}, written as follows,
\begin{align}\label{rad_aniso}
 X_{ra}&=X_{ca}\times\int\frac{d^3{\bf k}_5}{(2\pi)^32E_5}12g^2\frac{1}{k^2_\perp}\left(1+\frac{M_{HQ}^2}{s}e^{2\eta}\right)^{-2}\nonumber\\
    &\times\Big(1+ \hat{f}(E_5)\Big)\Theta_1(\tau-\tau_F)\Theta_2(E_p-E_5). 
\end{align}
Next, following the same description as in Eq. \eqref{trans_coll}, in the presence of anisotropy in the medium, the transport coefficient for the radiation processes can be decomposed as follows,
\begin{align}
\label{trans_rad}
 X_r = X_{r0} +X_{ra}.  
\end{align}

This analysis is carried under the approximation of soft gluon emission {\it i.e.,} $K_5\rightarrow 0$. The radiated gluons follow the Bose-Einstein phase space distribution, $\hat{f}(E_5)=\frac{1}{\exp{(\beta E_5)}-1}$. The emission of gluons from the HQs occurs under the two constraints; {the theta function}, $\Theta_2(E_p-E_5)$, ensures that the energy of emitted gluon ($E_5$) should not be greater than the HQs energy $(E_p)$. { In the dense medium, if the mean free time, the average time between two successive collisions, is of the order of formation time or larger, then the emission of gluon will be suppressed due to destructive interference, a phenomenon known as the Landau-Pomeranchuk-Migdal (LPM) suppression ~\cite{Gyulassy:1993hr,Klein:1998du}.  This has been taken care of in
a gross way through the theta function ,$\Theta_1(\tau-\tau_F)$, as done in Ref. ~\cite{Xu:2004mz, Mazumder:2013oaa,Das:2010tj}.
The invariant amplitude corresponding to radiative processes, $|\mathcal{M}|^2_{2\rightarrow 3}$ can be defined in terms of elastic invariant amplitude $|\mathcal{M}|^2_{2\rightarrow 2}$ \cite{Abir:2011jb}, given below, 

\begin{align}\label{matrixr}
    |\mathcal{M}|^2_{2\rightarrow 3}=|\mathcal{M}|^2_{2\rightarrow 2}\times 12g^2\frac{1}{k^2_\perp}\left(1+\frac{M_{HQ}^2}{s}e^{2\eta}\right)^{-2},
\end{align}
where $\eta$ is the rapidity of emitted gluons. The right-hand side of Eq.~\eqref{matrixr}, $\big(1+\frac{M_{HQ}^2}{s}e^{2\eta}\big)^{-2}$ represents the suppression factor because of the dead cone effect as detailed in Ref. \cite{ALICE:2021aqk,Dokshitzer:1991fd}.  While the HQs propagate through the medium, the energy loss may be suppressed due to the dead-cone effect. If the angle of emitted gluons with the direction of the HQs motion is smaller than $M_{HQ}/E_p$, the radiation of soft gluon suppresses ~\cite{Dokshitzer:2001zm}. The suppression factor depends on the mass of the HQs. 
The energy  and the transverse momentum of the emitted gluon  can be written in term of the rapidity variable as follow,
  \begin{align}\label{31}
    &E_5=k_\perp\cosh \eta, &&k_z=k_\perp\sinh \eta, 
\end{align} and,  $$d^3{ k}_5=d^2{ k}_\perp dk_z=2\pi k_\perp^2 dk_\perp \cosh\eta d\eta,$$ 
the interaction time and the interaction rate ($\Lambda$) are related because of restrictions on the $\tau_F$ as (for details we refer Ref. \cite{mustafa1998radiative,Xiang:2005ce}), 
\begin{align}\label{32}
    &\tau=\Lambda^{-1}>\tau_F=\frac{\cosh\eta}{k_\perp}.
\end{align}

Specifically one can write, $k_\perp>\Lambda\cosh\eta=(k_\perp)_{\text{min}}$, and $(k_\perp)_{\text{min}}$ denotes the minimum value of $k_{\perp}$. Further,  the second constraint on the emitted gluon energy, $E_5$,
\begin{align}\label{33}
   &E_p>E_5=k_\perp\cosh\eta,
  &&(k_\perp)_{\text{max}}=\frac{E_p}{\cosh\eta}.
\end{align}
Under the assumption of soft gluon emission, we keep $E_5=k_\perp \cosh \eta \ll T$.  
Now, the distribution function of the emitted gluon is approximated as follows,

\begin{align}\label{34}
 1+\hat{f}(E_5)\approx\frac{T}{k_\perp\cosh \eta}.   \end{align}

Finally, the contribution of radiative processes to the dynamics of the HQs in the anisotropic medium can be calculated through Eq.~\eqref{trans_rad}, while the elastic contribution can be calculated through Eq.~\eqref{trans_coll}. The overall contribution of both processes can be calculated by taking the summation of Eq.~\eqref{trans_coll} and Eq.~\eqref{trans_rad} as depicted in the following results section. We shall also discuss the potential impact of momentum anisotropy on the dynamics of the HQs by computing experimental observables, nuclear suppression factor, $R_{AA}$.

\begin{figure*}
 \centering
 \hspace{.15 cm}
 \subfloat{\includegraphics[scale=0.35]{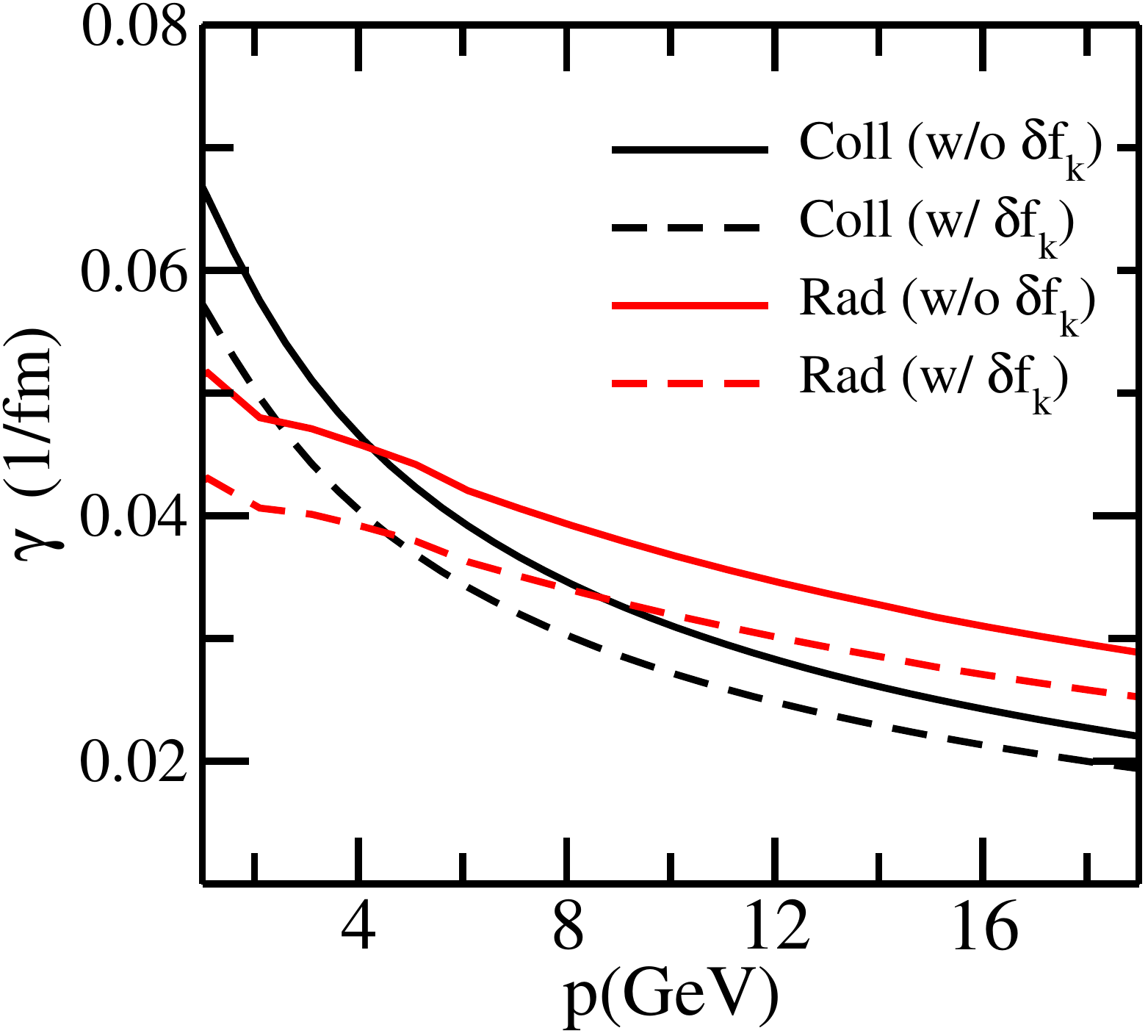}}
 \hspace{.15 cm}
 \subfloat{\includegraphics[scale=0.35]{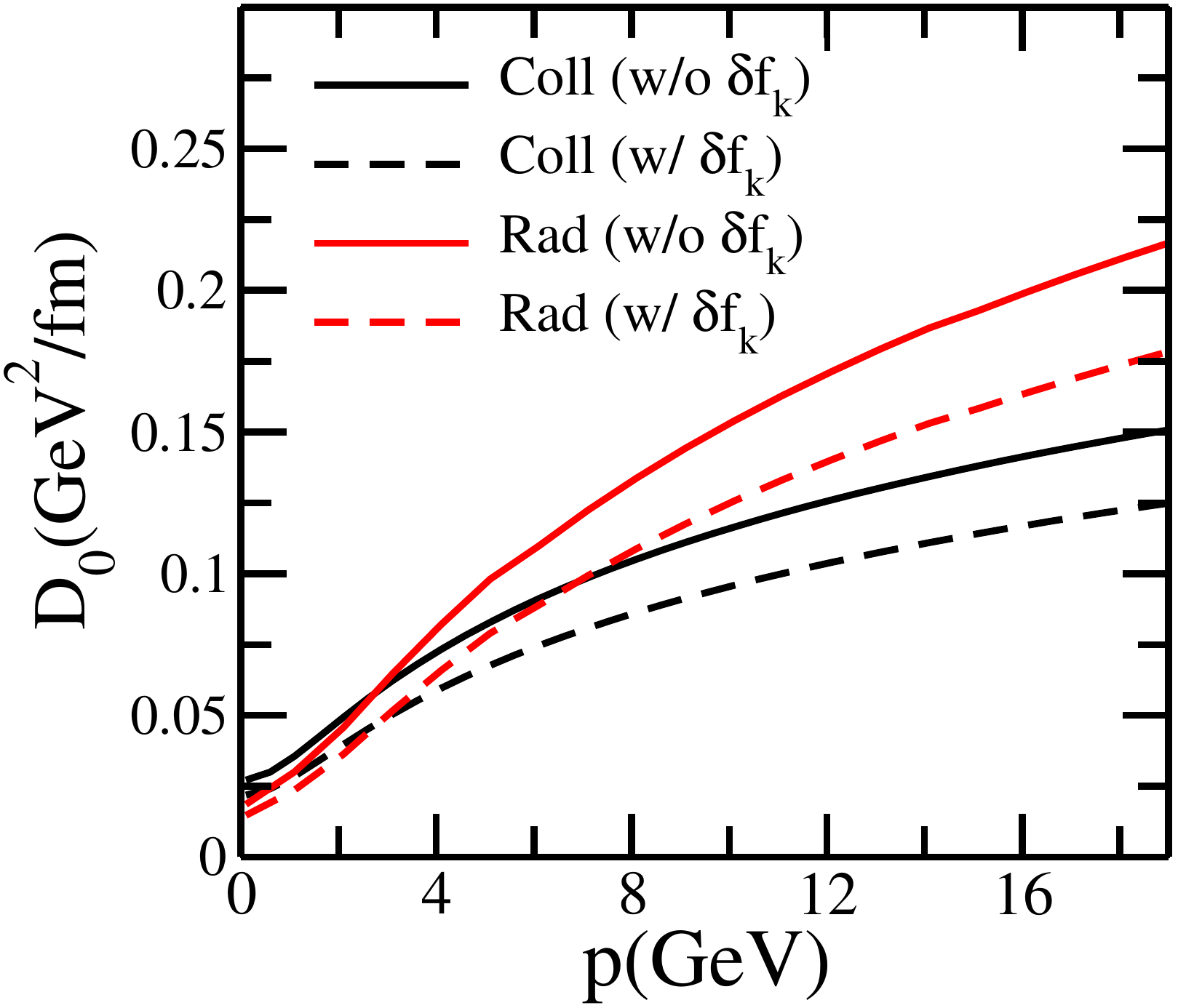}}
 \hspace{.15 cm}
 \subfloat{\includegraphics[scale=0.285]{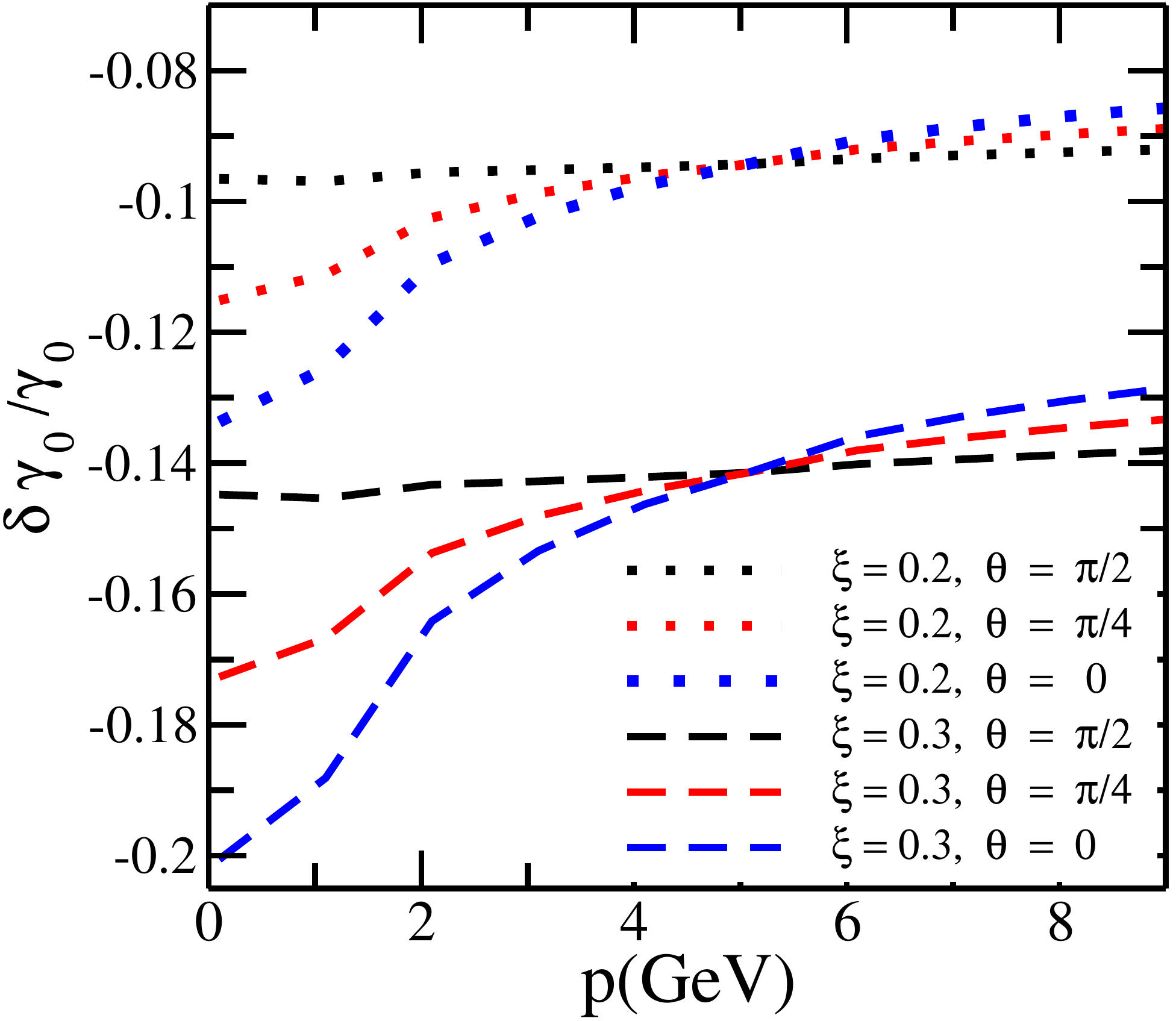}}
 \hspace{ 1.5 cm}
\caption{Momentum dependence of the HQs drag coefficient (left panel), diffusion coefficient  $D_0$ (middle) and anisotropic correction to isotropic drag coefficients as a function of momentum at a fixed temperature, $T = 360$ MeV  (right panel).}
\label{trans}
\end{figure*}

\begin{figure*}
 \centering
 \subfloat{\includegraphics[scale=0.31]{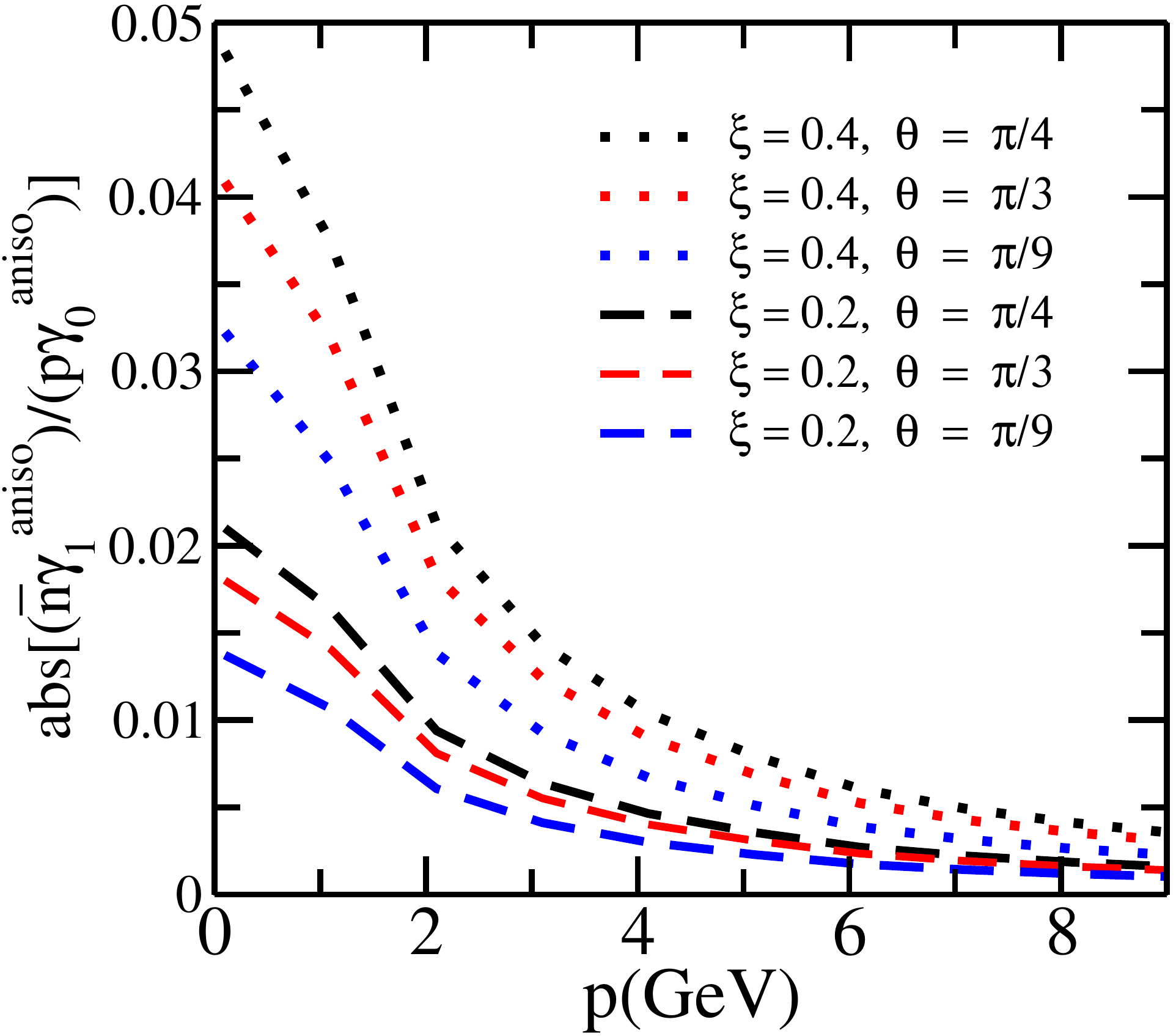}}
 \subfloat{\includegraphics[scale=0.34]{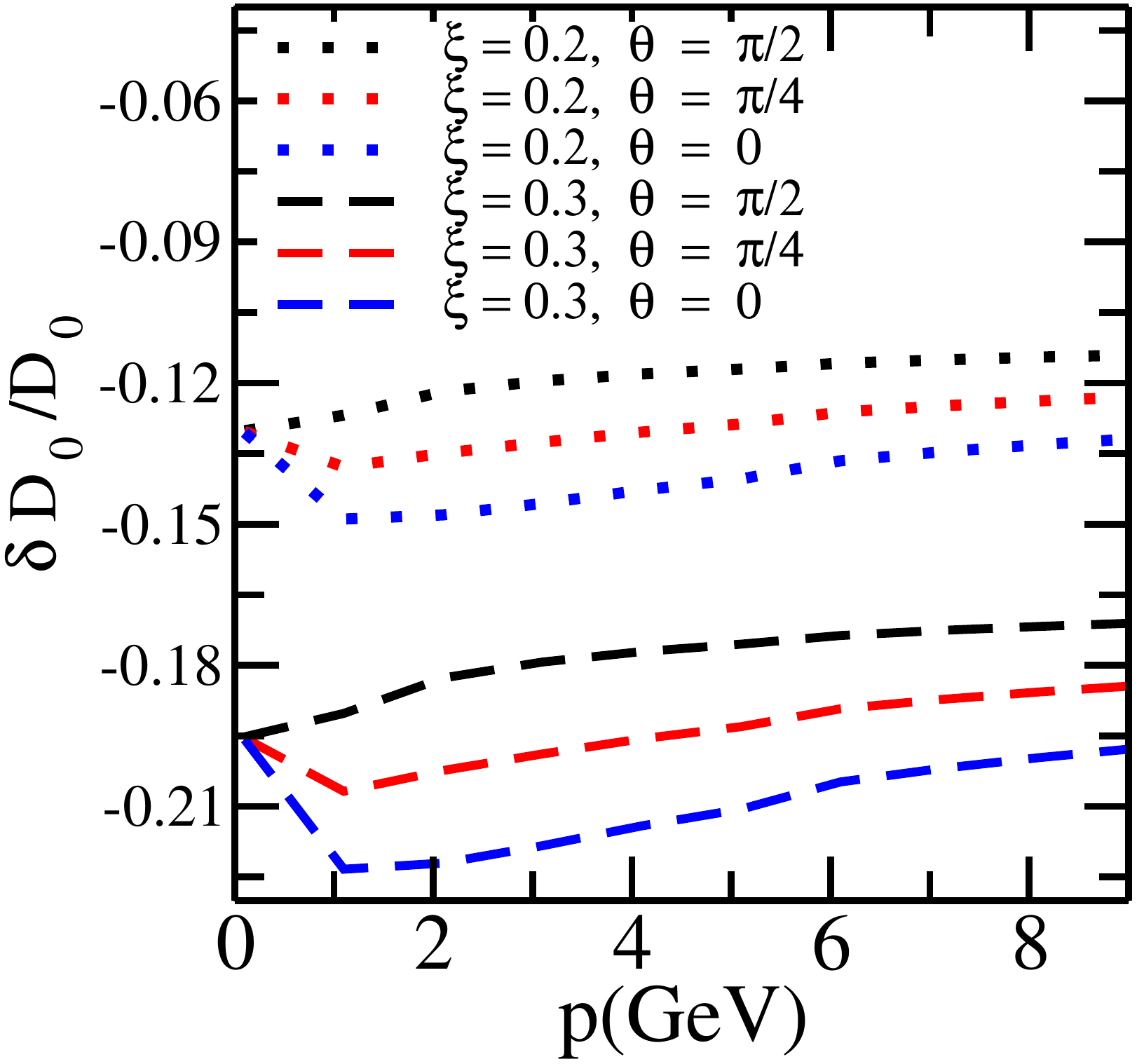}}
 \subfloat{\includegraphics[scale=0.35]{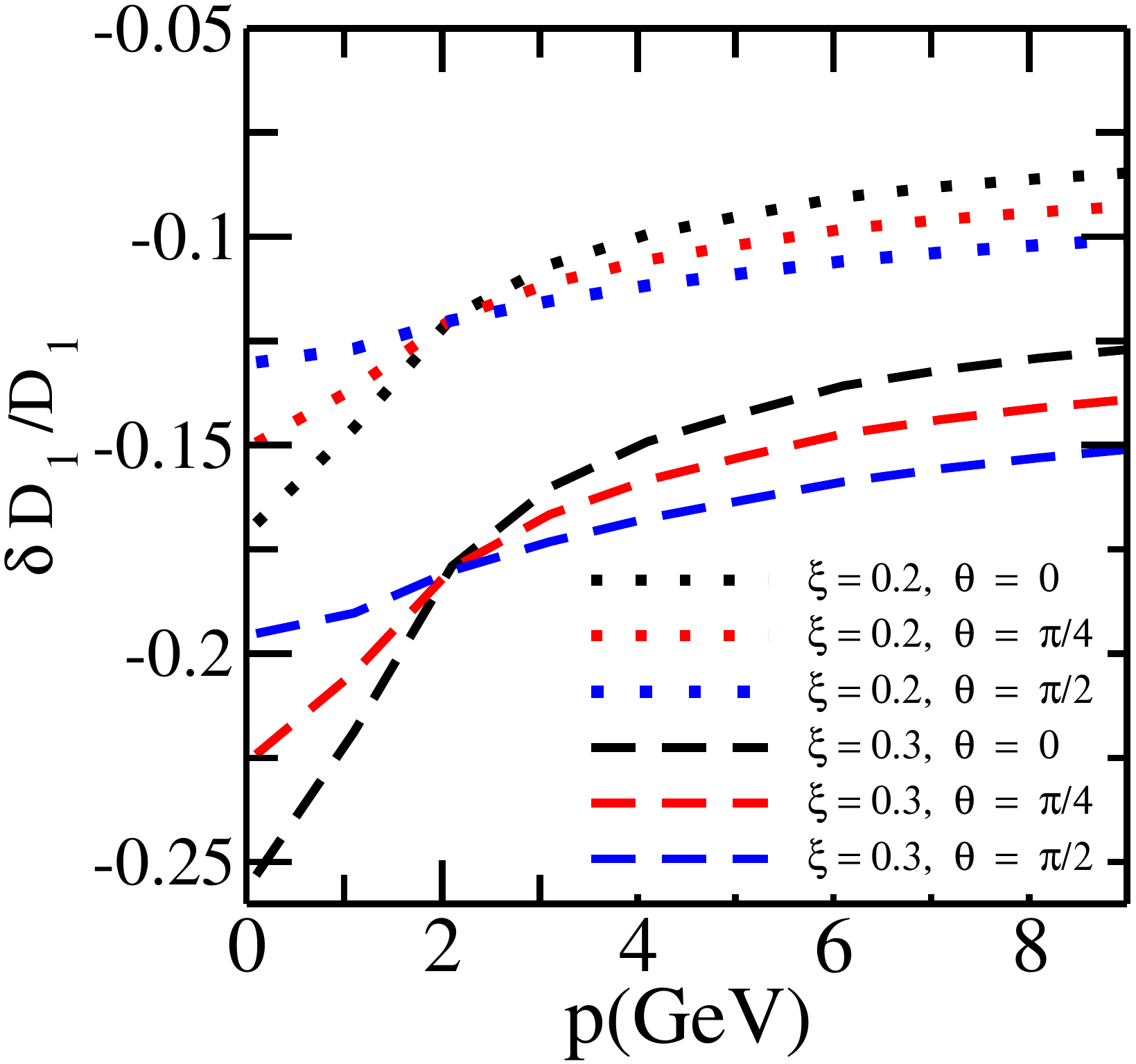}}

\caption{ Relative significance of $\gamma_1$ in comparison with $\gamma_0$ with a  momentum ($p$) at a fixed temperature, $T = 360$ MeV (left panel). Anisotropic corrections to $D_0$ (middle panel) and the anisotropic corrections to $D_1$ (right panel).}
\label{drag_correction}
\end{figure*}

\section{Results and Discussions}
\label{res}
We have obtained the transport coefficients of the HQ, namely charm quarks, in the presence of momentum anisotropy in the hot QCD medium. The mass of charm quark is taken as  1.3 GeV, and $\alpha_s$ is a two-loop running coupling constant \cite{PhysRevD.71.114510} are taken in the calculation. The effects of anisotropy enter through the momentum distribution of medium constituents that have been incorporated for the radiative and collisional processes. Since we assume the weak anisotropy limit, the anisotropic parameter is taken to be, $ \xi$= 0.2, 0.3, 0.4 ({ these  values of  $ \xi$  are in line with the numbers obtained within the hydrodynamics approach~\cite{Strickland:2014eua} }). The presence of anisotropy in the medium affects the transport coefficients of the HQ in the QGP medium and impacts the $R_{AA}$ substantially, as discussed below in detail.

\subsection{Transport coefficients in the anisotropic medium for elastic and inelastic process}
The propagation of the HQs (charm quarks) in the thermalized medium has been studied through the transport coefficient~\cite{ Das:2012ck}. This analysis found that the relative orientation of the HQs with the anisotropy vector plays a vital role in the HQ dynamics.
To incorporate that, we refer to Ref.~\cite{Kumar:2021goi} for collisional processes. 
This analysis explored the dynamics of the HQs in an anisotropic QCD medium for the collisional and radiation processes. We have noticed the effect of anisotropy due to the anisotropic angle. The transport coefficient of the HQ, namely, the drag coefficient, decomposes in two-component, $\gamma_0^{(\text{aniso})}$ and $\gamma_1^{(\text{aniso})}$ as defined in Eq.~\eqref{drag_0} and in Eq. \eqref{drag_1} respectively.

The effect of momentum anisotropy on the transport coefficients for the elastic and inelastic processes is depicted in Fig.~\ref{trans}.  The transport coefficients of the HQs are plotted as a function of momentum corresponding to the initial maximum temperature at RHIC energies, $T$ = 360 MeV. The effect of anisotropy has been calculated for $\theta = \pi/4$, and anisotropic strength, $\xi = 0.3$. The drag coefficients of the HQs are shown in Fig.\ref {trans} (left panel). 

 For both processes, we have observed that the HQs in the anisotropy medium suffer lesser hindrance while propagating through the QGP medium. The anisotropic contribution to the isotropic drag coefficient in Eq.~\eqref{drag_deco} suppresses the overall drag coefficient of the HQs in the anisotropic medium. { The effect of momentum anisotropy on the transport coefficients is almost similar for both elastic (represented by the black line) and inelastic processes (represented by the red line) processes}. The diffusion coefficient is plotted in Fig.~\ref {trans} (middle panel).
 For the radiative processes, the anisotropic correction to the isotropic drag coefficient strongly depends on the direction of anisotropy and anisotropic strength in the medium, especially at lower momentum regimes, as shown in Fig.~\ref {trans} (right panel). Also, the anisotropy direction alters at lower regimes of momentum than at high momentum. In Fig.~\ref{drag_correction} (left panel)  for the radiation processes, we found that the $\gamma_1^{(\text{aniso})}$ have negligible contribution at higher momentum. 
 We further realize that the additional drag component strongly depends on the strength of anisotropy.

We have shown the anisotropic correction to the HQs diffusion coefficient in Fig.~\ref{drag_correction} (middle and right panel), $D_0^{(\text{aniso})} = D_{0} + \delta D_0$, and $D_1^{(\text{aniso})} = D_{1} + \delta D_1$. The effect of momentum anisotropy is more noticeable at the low momentum regime. The HQs diffusion coefficient depends strongly on $\theta$ and $\xi$. One can realize that in the limit, $\xi\rightarrow{0}$,  anisotropic diffusion coefficient, $D_0^{(\text{aniso})}$ and $D_1^{(\text{aniso})}$ reduce to the isotropic diffusion coefficients one as described in Ref.~\cite{Svetitsky:1987gq} (for the same parameter).

\begin{figure*}
 \centering
 \hspace{.15 cm}
 \subfloat{\includegraphics[scale=0.31]{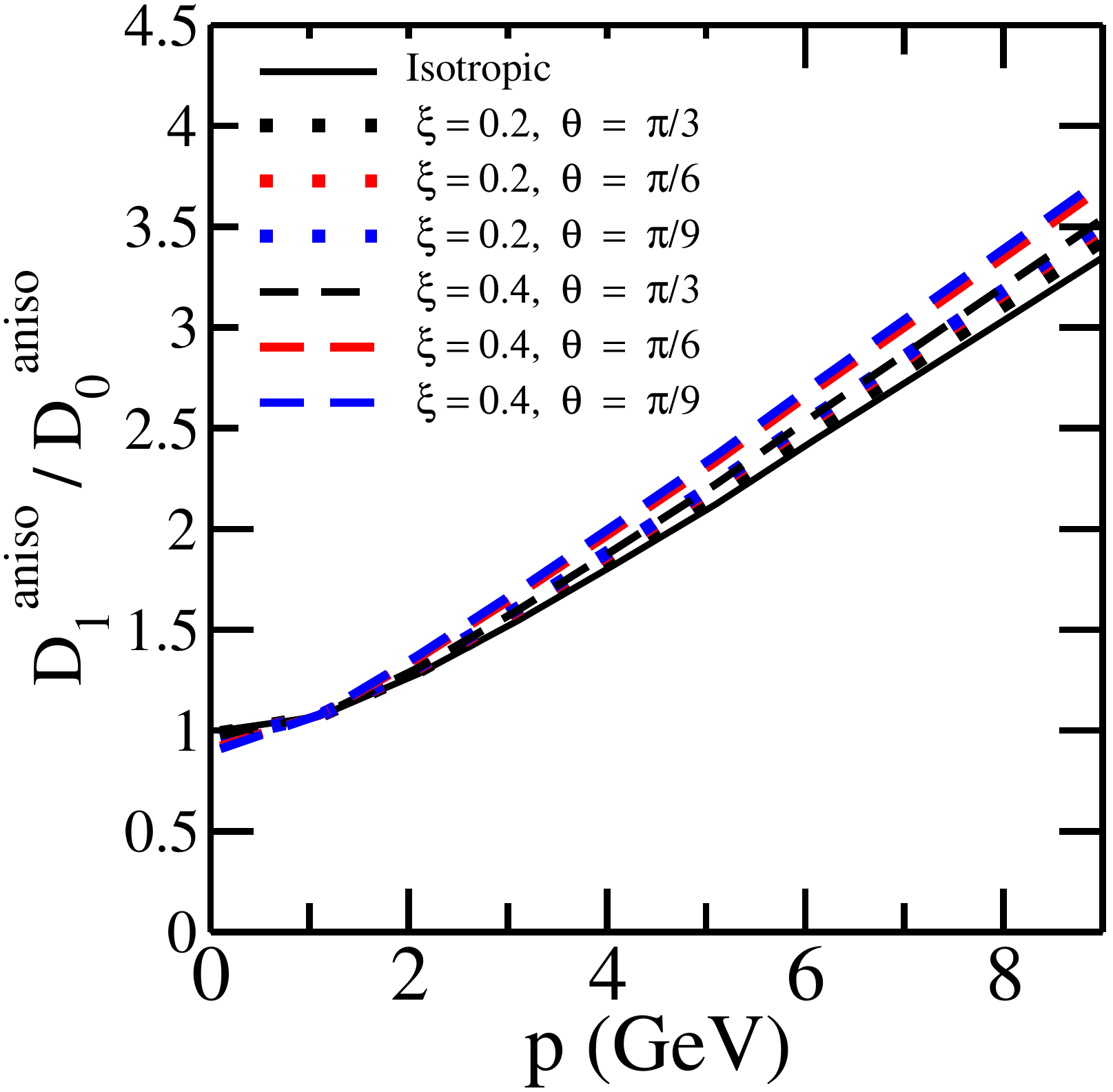}}
 \hspace{.15 cm}
 \subfloat{\includegraphics[scale=0.345]{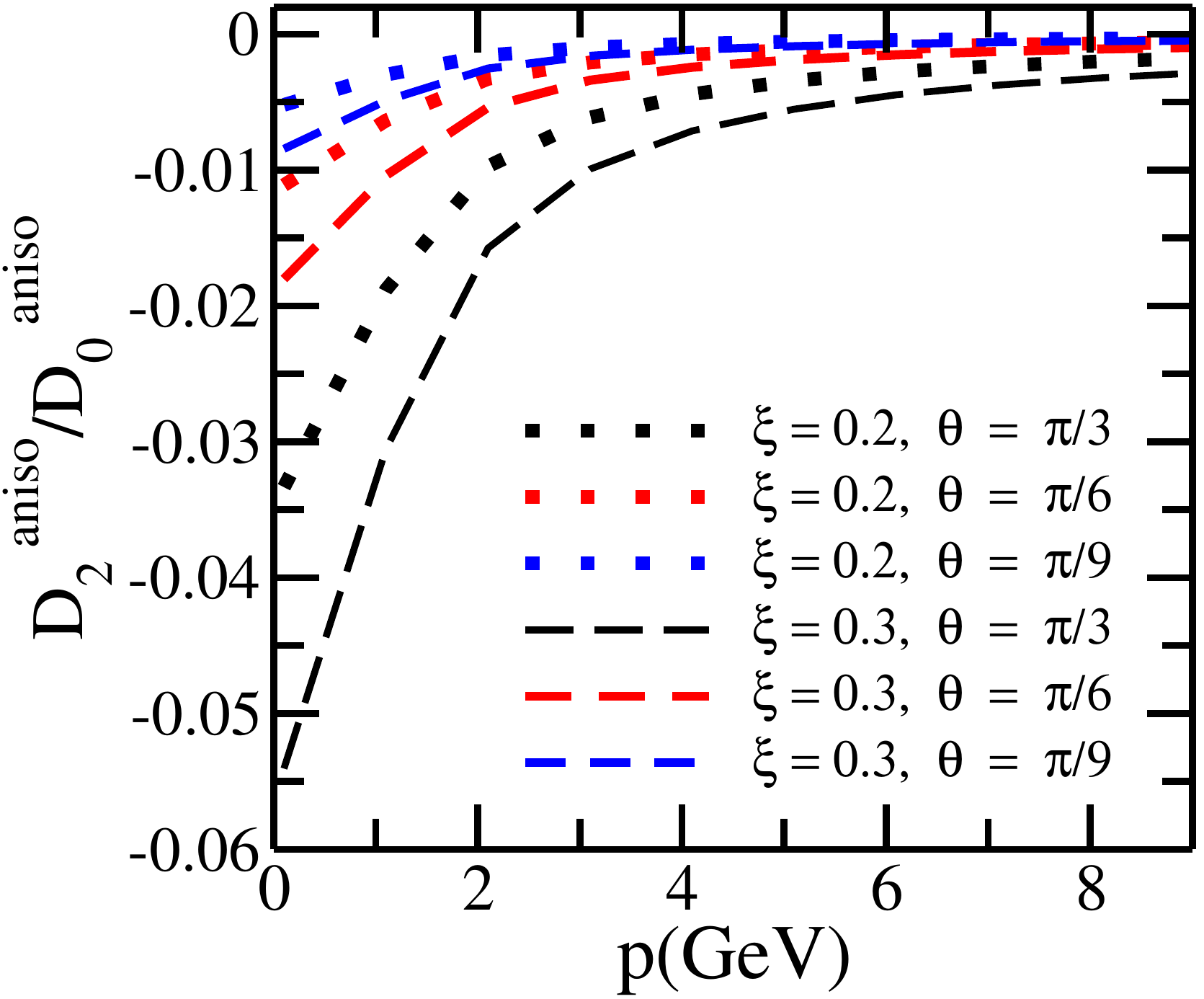}}
 \hspace{.15 cm}
\subfloat{\includegraphics[scale=0.34]{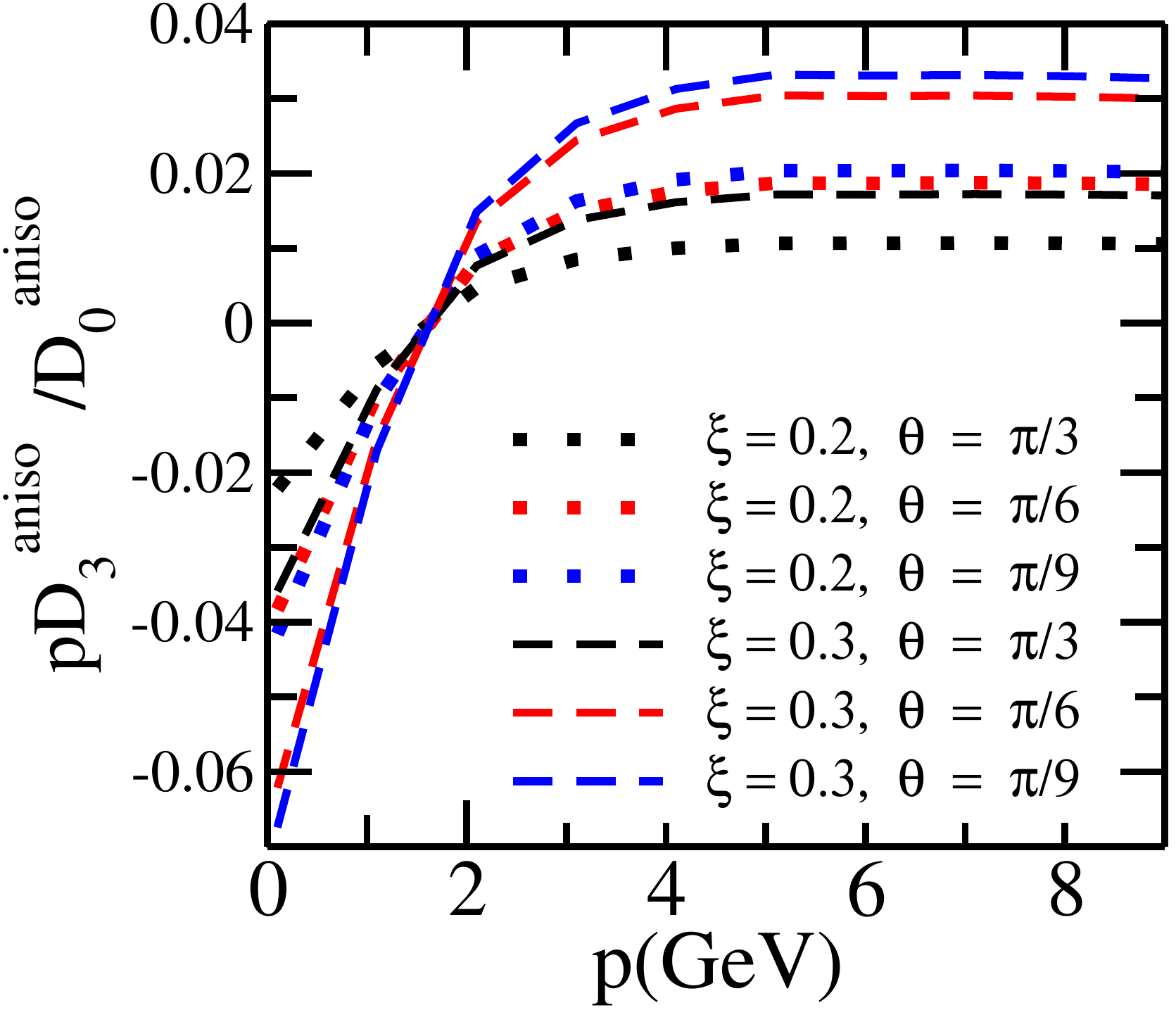}}
\caption{Relative Correction to the HQs diffusion coefficients for radiative processes: $D_1^{(\text{aniso})}/D_0^{(\text{aniso})}$ (left panel), $D_2^{(\text{aniso})}/D_0^{(\text{aniso})}$ (middle panel), and $pD_3^{(\text{aniso})}/D_0^{(\text{aniso})}$  (right panel) at $T = 360$ MeV.}
\label{diffusion_correction1}
\end{figure*}
The component, $D_1^{(\text{aniso})}$ and $D_0^{(\text{aniso})}$, are nearly equal in the limit $p\rightarrow 0$ as shown in Fig.~\ref {diffusion_correction1} (left panel). However, for the higher regimes of momentum, the component $D_1^{(\text{aniso})}$ is dominant over $D_0^{(\text{aniso})}$.
The other two components in Fig.~\ref {diffusion_correction1}, $D_2^{(\text{aniso})}$ (middle panel) and $D_3^{(\text{aniso})}$ (right panel) components of momentum diffusion show a strong dependence on the angle of anisotropy vector and the strength of anisotropy at lower momentum regimes. It should be noted that these additional coefficients become negligible as the system approaches isotropic medium, that is when the anisotropic parameter $\xi \rightarrow 0$.

\subsection{Nuclear suppression factor, $R_{AA}$}
\begin{figure*}
 \centering
 \subfloat{\includegraphics[scale=0.3]{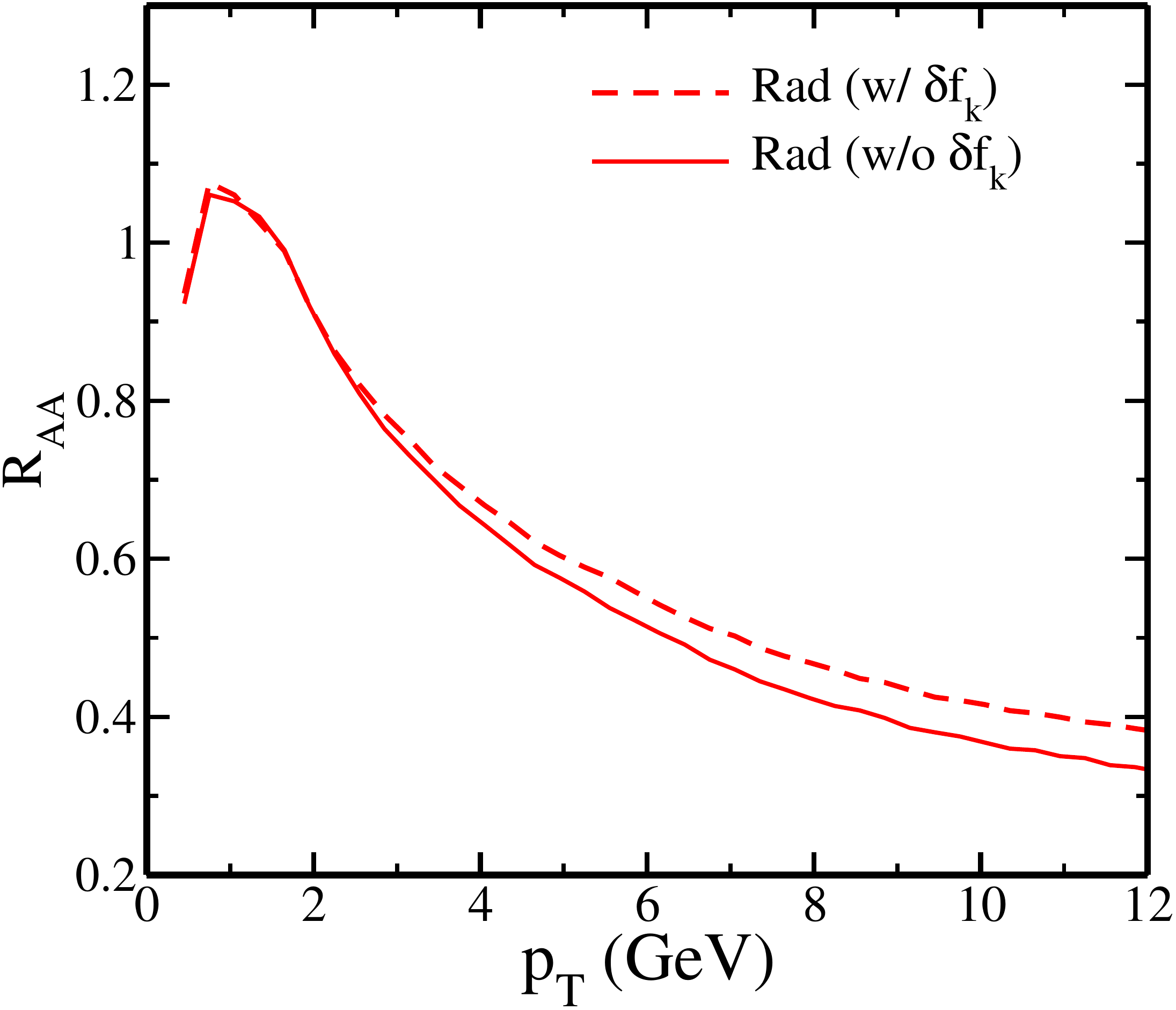}}
 \hspace{1 cm}
 \subfloat{\includegraphics[scale=0.29]{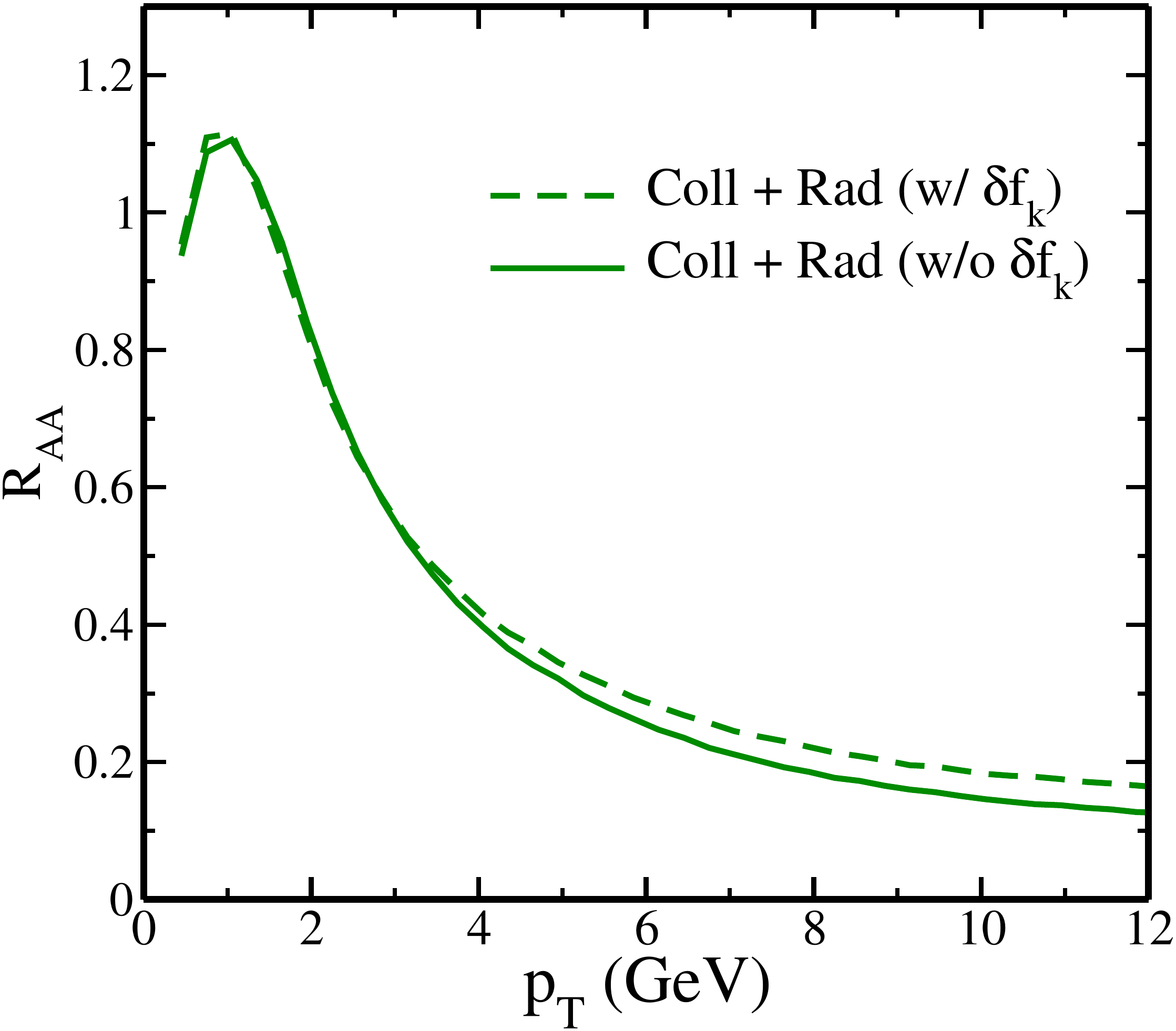}}

\caption{The $R_{AA}(p_T)$ of charm quarks  for  radiative processes (left panel) and summing both collisional and radiative processes (right panel) as a function of transverse momentum ($p_T$) at $T$  = $360$ MeV.}
\label{obervable}
\end{figure*}

To study the impact of the anisotropy on the experimental observable, we have estimated the nuclear suppression factor, $R_{AA}(p_T)$, in a static medium. The nuclear suppression factor, $R_{AA}(p_T)$ for the HQs is  defined as follows,
\begin{align}
 R_{AA}(p_T)=\frac{f_{\tau_f} (p_T )}{f_{\tau_0} (p_T)}.   
\end{align}

The final momenta spectrum of charm quarks is $f_{\tau_f} (p)$, at the end of time evolution  $\tau_f$, which is assumed 6 fm/c in our calculation. The initial momenta spectra, $f_{\tau_0}$, is taken according to the fixed order + next-to-leading log (FONLL) calculations ~\cite{Cacciari:2005rk,Cacciari:2012ny}. 
We employ stochastic Langevin dynamics ~\cite{Moore:2004tg}, as follows, 
\begin{align}
&dx_i=\frac{p_i}{E} dt,\\
& \label{lang}dp_i=-\gamma p_i\, dt+C_{ij}\rho_j\sqrt{dt},
\end{align}
where $dp_i$ and $dx_i$ are the change in momentum and position of the HQs, respectively, with the time step $dt$. In Eq.~\eqref{lang}, two forces act on the HQs motion in the hot QCD medium: the dissipative force and the stochastic force. The dissipative force governs the drag coefficient, $\gamma$, where $C_{ij}$ is the covariance matrix 
that describes stochastic force in terms of independent Gaussian-
normal distributed random variable, $\rho_j$, known as the white noise with $\langle\rho_i\rho_j\rangle=\delta_{ij}$ and $\langle\rho_j\rangle=0$. The covariance matrix is written as follows, 
\begin{align}
C_{ij} = \sqrt{2D_0}\left(\delta_{ij}-\frac{p_ip_j}{p^2}\right)+\sqrt{2D_1}\frac{p_ip_j}{p^2}.
\end{align}
Thus, at $p \rightarrow 0$,  the coefficient, $D_0=D_1=D$, further $C_{ij}=\sqrt{2D}\delta_{ij}$, where $D$ is the diffusion coefficient of the HQ. We take transport coefficients as an input kernel in the Langevin equation to compute the momentum evolution of the HQs in the hot QCD medium.

We investigate the effect of momentum anisotropy on $R_{AA}$ of the HQs considering both collisional and radiative processes. In Fig.~\ref{obervable}, $R_{AA}$ is plotted as a function of $p_T$ at $T= 360$ MeV for radiation (left panel), and summing both collisional and radiative processes (right panel). The effect of momentum anisotropy has been calculated for the anisotropic angle, $\theta = \pi/9$, and anisotropic strength, $\xi = 0.4$. The emission of gluons by the HQs in the QGP medium significantly changes the value of $R_{AA}$. The impact of momentum anisotropy is quite visible on $R_{AA}$. In the higher $p_T$ regimes, the momentum anisotropy increases $R_{AA}$, approximately 18$\%$ (indicated by the red dashed line), leading to less suppression. This could be a consequence of the momentum anisotropy delivering a lesser hindrance for the charm quark motion in the QGP at high $p_T$. In Fig.~\ref{obervable} (right panel) we have included both collisional and radiative contributions showing a stronger suppression. One can say that the energy loss of the HQs gets suppressed in the presence of anisotropy in the hot QCD medium.

\section{Conclusions and Outlook}
\label{con}
We have studied the transport coefficients of the HQ in the anisotropic hot QCD medium within the framework of Fokker-Plank dynamics while considering both collisional and radiative processes. The analysis has been done in a weakly anisotropic medium where the anisotropic aspects enter through the non-equilibrium distribution of the medium constituents. We have considered a general tensor decomposition for the transport coefficients (the drag coefficient split into two components and the momentum diffusion tensor of the HQs split into four components). We observed that the correction term (because of momentum anisotropy) of the transport coefficient significantly contributes to the isotropic components of the HQs transport coefficients.  Possibly, strong anisotropy may affect the additional component of transport coefficients. Our primary emphasis lies in studying radiation processes; through our study, we have come to understand that the anisotropic effects within the medium are highly reliant on the orientation of the HQs motion with respect to the direction of anisotropy. 

A significant influence of the anisotropy on the drag coefficient and diffusion coefficients of the HQs has been noticed. In particular, the impact of anisotropy is seen to be more noticeable at lower momentum regimes. This suggests that the average momentum transfer experienced by the HQs is influenced by both the direction and strength of momentum anisotropy within the medium. Therefore, the inclusion of anisotropy seems to be crucial for the phenomenological consistency of the HQs motion in the QGP medium.

We further expand the current analysis while considering the physical observables, {\it i.e.,}  $R_{AA}$ of the HQs in the QGP medium. We have solved the Langevin equation, where the HQs move under the dissipative and random force. The transport coefficients, namely, the drag and the diffusion, have served as the input to the Langevin dynamics.  Within the anisotropic medium, we have observed a decrease in the magnitude of the transport coefficients of the HQs within the selected momentum regimes. This reduction subsequently leads to less suppression in the $R_{AA}$ for the collisional and radiative processes. The energy loss of the HQs is found to be suppressed in the presence of momentum anisotropy in the medium. 
{Since momentum anisotropy reduces the drag coefficient of heavy quarks inside the
medium, it can reduce the $v_2$. However, the angle dependence can also enhance the $v_2$. It will be interesting to study heavy quark $v_2$ in an anisotropic medium which is beyond the scope of the present work and will be addressed in the near future.}

To fully investigate the phenomenological aspects of the HQs at RHIC and the LHC energies, it is crucial to systematically consider the scattering process (matrix element) of the HQs within  {(3 + 1)D} expansion of the QGP medium. Such an approach ensures a thorough analysis that accounts for both the specific characteristics of the medium and the dynamics of the HQs. Exploring the dynamics of the HQs within a strong anisotropic medium presents an intriguing effort. We anticipate that the inclusion of the additional drag and diffusion coefficients arising from the anisotropy within the medium might play a significant role in such a strongly anisotropic medium. These aspects will be a matter of investigation in the immediate future.

\section{Acknowledgements} 
J. P. acknowledges Manu Kurian for useful discussions. V. C. and S. K. D. acknowledge the SERB Core Research Grant (CRG) [CRG/2020/002320]. S.K.D. acknowledges the support from DAE-BRNS, India, Project No.: 57/14/02/2021-BRNS.

\appendix

 \section{{Anisotropy vector projections in  center-of-mass frame}}
\label{apendix}
\begin{widetext}

\begin{align}
\left(\textbf{v}_{cm}\hspace{0.05cm} \mathord{\cdot} \hspace{0.05cm} \Hat{\textbf{p}}_{cm}\right) &=\gamma_{cm}\left[\frac{(p^2+pq\cos\chi)}{E_p+E_q}-v^2_{cm}E_p\right],\\
  \label{A2}N^2 &= v_{cm}^2-\frac{\left(\textbf{v}_{cm}\hspace{0.05cm} \mathord{\cdot} \hspace{0.05cm} 
  \Hat{\textbf{p}}_{cm}\right)^2}{\Hat{p}^2_{cm}}, 
     \end{align}

 \begin{align}
  v_{cm}^2 &= \frac{(p^2+q^2+2pq\cos\chi)}{(E_p+E_q)^2},
     \end{align}
 \begin{align}
 \label{A4}\left(\Hat{\textbf{x}}_{cm}\hspace{0.05cm} \mathord{\cdot} \hspace{0.05cm} \textbf{n}\right) &= \frac{\gamma_{cm}}{\Hat{p}_{cm}}\left[p\cos\theta-E_p\frac{(p\cos\theta + q\cos\chi\cos\theta +q\sin\chi\cos\phi\sin\theta)}{E_p+E_q}\right],
 \\ \nonumber
 \left(\Hat{\textbf{y}}_{cm}\hspace{0.05cm} \mathord{\cdot} \hspace{0.05cm} \textbf{n}\right) &=  N^{-1}\left[\frac{(p\cos\theta + q\cos\chi\cos\theta +q\sin\chi\cos\phi\sin\theta)}{E_p+E_q}\right]
  \\ 
 &- \left(\textbf{v}_{cm}\hspace{0.05cm} \mathord{\cdot} \hspace{0.05cm} \Hat{\textbf{p}}_{cm}\right)\frac{\gamma_{cm}}{\Hat{p}_{cm}}\left[p\cos\theta-E_p\frac{(p\cos\theta + q\cos\chi\cos\theta +q\sin\chi\cos\phi\sin\theta)}{E_p+E_q}\right],\\
 \left(\Hat{\textbf{z}}_{cm}\hspace{0.05cm} \mathord{\cdot} \hspace{0.05cm} \textbf{n}\right) &= \gamma_{cm}N^{-1}\frac{1}{\Hat{p}_{cm}(E_p+E_q)}pq\sin\chi\sin\phi\sin\theta,
    \end{align}
 \begin{align}
 (\Hat{\textbf{x}}_{cm}\hspace{0.05cm} \mathord{\cdot} \hspace{0.05cm} \textbf{p}) &= \frac{\gamma_{cm}}{\Hat{p}_{cm}}\left[p^2-E_p\frac{(p^2+pq\cos\chi)}{(E_p+E_q)}\right],\\
 \label{A8}(\Hat{\textbf{y}}_{cm}\hspace{0.05cm} \mathord{\cdot} \hspace{0.05cm} \textbf{p}) &= N^{-1}\left[\frac{(p^2+pq\cos\chi)}{(E_p+E_q)}-\left(\textbf{v}_{cm}\hspace{0.05cm} \mathord{\cdot} \hspace{0.05cm} \Hat{\textbf{p}}_{cm}\right) \frac{\gamma_{cm}}{\Hat{p}^2_{cm}}
   \left(p^2-E_p\frac{(p^2+pq\cos\chi)}{(E_p+E_q)}\right)\right].
 \end{align}
\end{widetext}


\bibliography{ref1}
\end{document}